\documentclass[aps, prd, floatfix, nofootinbib, superscriptaddress, twocolumn]{revtex4-1}

\usepackage{latexsym}
\usepackage{amsmath}
\usepackage{amssymb}
\usepackage{amsfonts}
\usepackage{textcomp}
\usepackage{color}
\usepackage{CJKutf8}

\usepackage[mathscr,scaled=1.15]{urwchancal}
\DeclareFontFamily{OT1}{pzc}{}
\DeclareFontShape{OT1}{pzc}{m}{it}%
{<-> s * [1.15] pzcmi7t}{}
\DeclareMathAlphabet{\mathpzc}{OT1}{pzc}{m}{it}

\usepackage{color}

\usepackage{supertabular}
\usepackage{placeins}
\usepackage{epsfig}
\usepackage{graphicx}

\definecolor{purple}{rgb}{0.5,0,0.5}
\definecolor{blue}{rgb}{0.0,0,0.9}
\definecolor{prdblue}{rgb}{0.133,0.118,0.498}
\usepackage[colorlinks=true, pdfstartview=FitV, linkcolor=prdblue, citecolor= prdblue, urlcolor=prdblue]{hyperref}

\makeatletter

\setbox0\hbox{$\xdef\scriptratio{\strip@pt\dimexpr
    \numexpr(\sf@size*65536)/\f@size sp}$}

\newcommand{\scriptveryshortarrow}[1][3pt]{{%
    \hbox{\rule[\scriptratio\dimexpr\fontdimen22\textfont2-.2pt\relax]
               {\scriptratio\dimexpr#1\relax}{\scriptratio\dimexpr.4pt\relax}}%
   \mkern-4mu\hbox{\let\f@size\sf@size\usefont{U}{lasy}{m}{n}\symbol{41}}}}

\makeatother

\begin{document}

\begin{CJK}{UTF8}{song}

\title{$\,$\\[-6ex]\hspace*{\fill}{\normalsize{\sf\emph{Preprint no}.\ NJU-INP 049/21}}\\[1ex]
Valence quark ratio in the proton}


\date{2022 May 26}

\author{Zhu-Fang Cui} 
\affiliation{School of Physics, Nanjing University, Nanjing, Jiangsu 210093, China}
\affiliation{Institute for Nonperturbative Physics, Nanjing University, Nanjing, Jiangsu 210093, China}
\author{Fei Gao} 
\affiliation{
Centre for High Energy Physics, Peking University, Beijing 100871, China}
\author{Daniele Binosi}
\email{binosi@ectstar.eu}
\affiliation{European Centre for Theoretical Studies in Nuclear Physics
and Related Areas, Villa Tambosi, Strada delle Tabarelle 286, I-38123 Villazzano (TN), Italy}
\author{\\Lei Chang} 
\email[]{leichang@nankai.edu.cn}
\affiliation{School of Physics, Nankai University, Tianjin 300071, China}
\author{Craig D.~Roberts}
\email[]{cdroberts@nju.edu.cn}
\affiliation{School of Physics, Nanjing University, Nanjing, Jiangsu 210093, China}
\affiliation{Institute for Nonperturbative Physics, Nanjing University, Nanjing, Jiangsu 210093, China}
\author{Sebastian M.~Schmidt}
\affiliation{Helmholtz-Zentrum Dresden-Rossendorf, Dresden D-01314, Germany}
\affiliation{RWTH Aachen University, III. Physikalisches Institut B, Aachen D-52074, Germany}

\begin{abstract}
Beginning with precise data on the ratio of structure functions in deep inelastic scattering (DIS) from $^3$He and $^3$H, collected on the domain $0.19 \leq x_B \leq 0.83$, where $x_B$ is the Bjorken scaling variable, we employ a robust method for extrapolating such data to arrive at a model-independent result for the $x_B=1$ value of the ratio of neutron and proton structure functions.  Combining this with information obtained in analyses of DIS from nuclei, corrected for target-structure dependence, we arrive at a prediction for the proton valence-quark ratio: $\left. d_v/u_v \right|_{x_B\to 1} = 0.230 (57)$.  Requiring consistency with this result presents a challenge to many descriptions of proton structure.
\end{abstract}

\maketitle

\end{CJK}


\noindent\emph{1.$\;$Introduction} ---
In the Standard Model of particle physics, there are two flavours of light quark -- up ($u$) and down $(d)$ quarks; and the proton is comprised of two valence $u$ quarks and one valence $d$ quark, which are bound by the exchange of gluons, \emph{viz}.\ the gauge bosons of quantum chromodynamics (QCD).  Such valence-quarks are not the constituent quarks \cite{GellMann:1964nj, Zweig:1981pd} used to build quantum mechanics descriptions of the properties of baryons (neutrons, protons, etc.) \cite{Lucha:1991vn, Capstick:2000qj, Giannini:2015zia}, although links can be drawn between them \cite[Sec.\,2.3]{Roberts:2021nhw}.  Moreover, quantum field theory is far more complex than quantum mechanics, whilst preserving all its basic features.  These things together mean that any attempt to express the proton wave function in terms of QCD's Lagrangian degrees-of-freedom will require infinitely many terms \cite{Brodsky:1989pv}, each one containing different numbers of gluons, quarks and antiquarks.  Approached this way, the problem of solving for the proton wave function has thus far been intractable.

Notwithstanding the challenges involved, the proton wave function is a very desirable quantity.  Indeed, analogous to the wave function for the hydrogen atom, a sound result for the proton wave function, $\Psi_p$, would enable many basic features of QCD to be understood.  For instance, even though it is impossible to speak of their precise number, with $\Psi_p$ in hand one can compute matrix elements that express the \emph{number density distributions} within the proton of QCD's gluons and quarks, \emph{i.e}., the proton parton distribution functions (DFs) \cite[Sec.\,4]{Ellis:1991qj}.

\begin{figure}[b]
\centerline{%
\includegraphics[clip, width=0.40\textwidth]{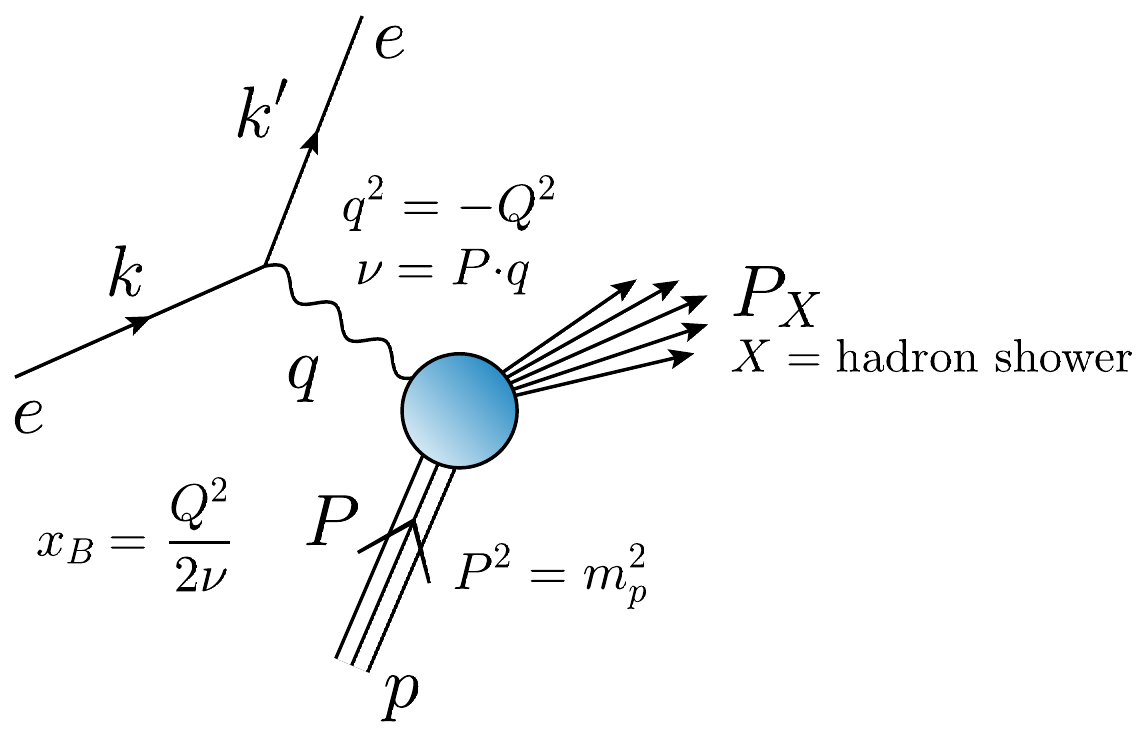}}
\caption{\label{Fdis}
electron+proton ($ep$) scattering, with the initial-state proton disintegrated by collision with the virtual photon, producing a cascade, $X$, of final state hadrons.  With $m_p$ being the proton mass, the Bjorken limit is defined by the following kinematics: $Q^2 \gg m_p^2$, $\nu\gg m_p^2$, $x_B\,=$\,fixed ($0<x_B<1$).
}
\end{figure}

The proton DFs have long been the subject of intense study \cite{Holt:2010vj, Rojo:2015acz, Peng:2016ebs, Hen:2016kwk}.  They can be inferred from measurements using, \emph{e.g}., the deep inelastic scattering (DIS) process drawn in Fig.\,\ref{Fdis}, when the kinematic variables satisfy the Bjorken conditions \cite{Bjorken:1968dy}.  It was using just such processes that quarks were discovered \cite{Taylor:1991ew, Friedman:1991nq, Kendall:1991np, Friedman:1991ip}.

Returning to the issue of valence quarks, given that the proton is that object which appears as the lowest-mass bound-state in the scattering of two $u$-quarks and one $d$ quark, it is natural to ask for the ratio of $u$ and $d$ quark number densities in this system.
Importantly, the value of this ratio on the far-valence domain, $x_B \simeq 1$, is one of the keenest available discriminators between competing descriptions of proton structure \cite{Roberts:2013mja}; and neutron $(n)$ structure, too, because SU$(2)$-flavour (isospin) symmetry is an accurate approximation for strong interactions.

Before considering possible answers, it is first necessary to observe that the simplest description of high-energy scattering processes is obtained by formulating the problem using light-front coordinates, in which case one writes the proton's four-momentum $P=(P_0,\vec{P}_\perp = (P_1,P_2),P_3) \to (P_-,\vec{P}_\perp,P_+)$ with $P_- = [|\vec{P}_\perp|^2 +m_p^2]/P_+$ \cite{Dirac:1949cp}.  Formulated in these terms, then at leading order in analyses using perturbative QCD, the Bjorken variable, $x_B$ in Fig.\,\ref{Fdis}, can be identified with the light-front fraction of the proton's momentum carried by the parton which participates in the scattering process \cite{Kogut:1969xa, Bjorken:1970ah}: $x = l_+/P_+$.  The proton DFs are then accessible, \emph{e.g}., via projection of the proton's Poincar\'e covariant wave function onto the light-front.

Maintaining Poincar\'e covariance in calculations of the proton wave function and DIS observables is crucial.  Yet, results based on nonrelativistic quark model wave functions are still cited.  The simplest of these are based on few-body quantum mechanics, with two-body potentials between constituent-quarks, and realise a SU$(2)$ spin-flavour symmetry in the proton wave function.  In this case, the proton wave function is expressed in terms of $U$ and $D$ constituent-quarks, which are largely uncorrelated, so that their individual wave functions are proportional; consequently, the $D/U$ ratio is $1/2$ for all values of $x$.  This prediction was largely rendered untenable by the first experiments to provide data that could be interpreted in terms of the valence-quark $d/u$ ratio \cite{Bodek:1973dy, Poucher:1973rg}.
Herein, we use data from the most recent such experiment \cite{Abrams:2021xum} -- MARATHON -- to deliver a model-independent prediction for the ratio of proton valence-quark densities.

\medskip

\noindent\emph{2.$\;$Structure function ratio: theory} ---
On any domain for which the Bjorken conditions are satisfied, the ratio of valence-quark DFs can be obtained by separately considering DIS reactions from neutron and proton targets and forming a particular ratio of Poincar\'e-invariant structure functions that appear in the expressions for the two cross-sections \cite[Eq.\,(2.45)]{Holt:2010vj}:
\begin{align}
\label{F2nF2p}
\frac{F_2^n(x)}{F_2^p(x)} = \frac{u_v(x) + 4 d_v(x) + 6 d_s(x) + \Sigma(x)}{4 u_v(x)+d_v(x) + 6 u_s(x)+\Sigma(x)}\,,
\end{align}
where $u_v(x)$, $d_v(x)$ are the proton valence-quark DFs; $u_s(x)$, $d_s(x)$, the associated sea-quark DFs; and $\Sigma(x)$ is the remaining flavour-symmetric sea-quark contribution.  In explanation, sea-quarks are generated by interactions between the proton's valence-quarks and are a necessary part of the proton wave function when it is expressed in terms of QCD's Lagrangian degrees-of-freedom.

It is now known \cite{SeaQuest:2021zxb} that $d_s(x)/u_s(x)\neq 1$, deviating by possibly as much as a factor of $1.5$ on $0.1<x<0.4$.  Nevertheless, in comparison with valence DFs, all sea distributions in Eq.\,\eqref{F2nF2p} are negligible on $x\gtrsim 0.2$ (see, \emph{e.g}., Ref.\,\cite[Ch.\,18]{Zyla:2020zbs} and the discussions in Refs.\,\cite{Chang:2021utv, Chang:2022jri}); hence:
\begin{align}
\label{F2nF2pV}
\frac{F_2^n(x)}{F_2^p(x)} \stackrel{x\gtrsim 0.2}{=} \frac{1 + 4 d_v(x)/u_v(x) }{4 +d_v(x)/u_v(x)}\,.
\end{align}
The limiting cases $d_v(x)\equiv 0$ and $u_v(x)\equiv 0$ yield the Nachtmann bounds \cite{Nachtmann:1973mr}:
\begin{equation}
\label{F2nF2pVNachtmann}
1/4 \leq F_2^n(x)/F_2^p(x) \leq 4\,.
\end{equation}

Owing to scaling violations \cite[Sec.\,4]{Ellis:1991qj}, $F_2^n(x)/F_2^p(x)$ actually depends on the $Q^2$ value at which it is measured; except at the endpoint $x=1$, whereat the ratio is a fixed point under QCD evolution \cite{Holt:2010vj}.  This feature, which can be verified by adapting the analysis in Ref.\,\cite[App.\,A]{Cui:2021sfu}, is what empowers the ratio $d_v(x)/u_v(x)$ as an acute discriminator between pictures of proton structure.

\medskip

\noindent\emph{3.$\;$Structure function ratio: experiment} ---
The proton is stable and has been used as a target in high-energy experiments for almost one hundred years \cite{Cockroft:1932I}.  Isolated neutrons, however, decay rather quickly; so the big issue in obtaining reliable measurements of $F_2^n(x)/F_2^p(x)$ is centred on the construction of a ``free neutron'' target.  Beginning with Refs.\,\cite{Bodek:1973dy, Poucher:1973rg}, many experiments have used the deuteron as the path to a neutron target; but even though this is a weakly bound system, the characterisation of proton-neutron interactions introduces a large theory uncertainty in the extracted ratio on $x\gtrsim 0.7$ \cite{Whitlow:1991uw}.

A different approach is to perform DIS measurements on $^3$He and $^3$H and then take the ratio of scattering rates from these two nuclei.  In this case, nuclear interaction effects largely cancel when extracting $F_2^n(x)/F_2^p(x)$ \cite{Afnan:2000uh, Pace:2001cm}.  The challenge here is developing a method for handling the radioactive $^3$H target; but after years of development, such an experiment was recently completed \cite{Abrams:2021xum}.  Adding credibility to these new results, the data obtained is, within mutual uncertainties, in agreement with an analysis of nuclear DIS reactions, on targets ranging from the deuteron to lead, which accounts for the impact of short-range correlations in the nuclei \cite{Segarra:2019gbp}.

The possibility that MARATHON data omit a systematic effect deriving from model-dependent assumptions about nuclear structure is raised in Refs.\,\cite{Segarra:2021exb, Cocuzza:2021rfn}.  More analyses are required before this hypothesis can be validated and its potential impact reliably quantified.  We therefore proceed with analysis of the data reported in Ref.\,\cite{Abrams:2021xum}.

\begin{figure}[t]
\centerline{%
\includegraphics[clip, width=0.43\textwidth]{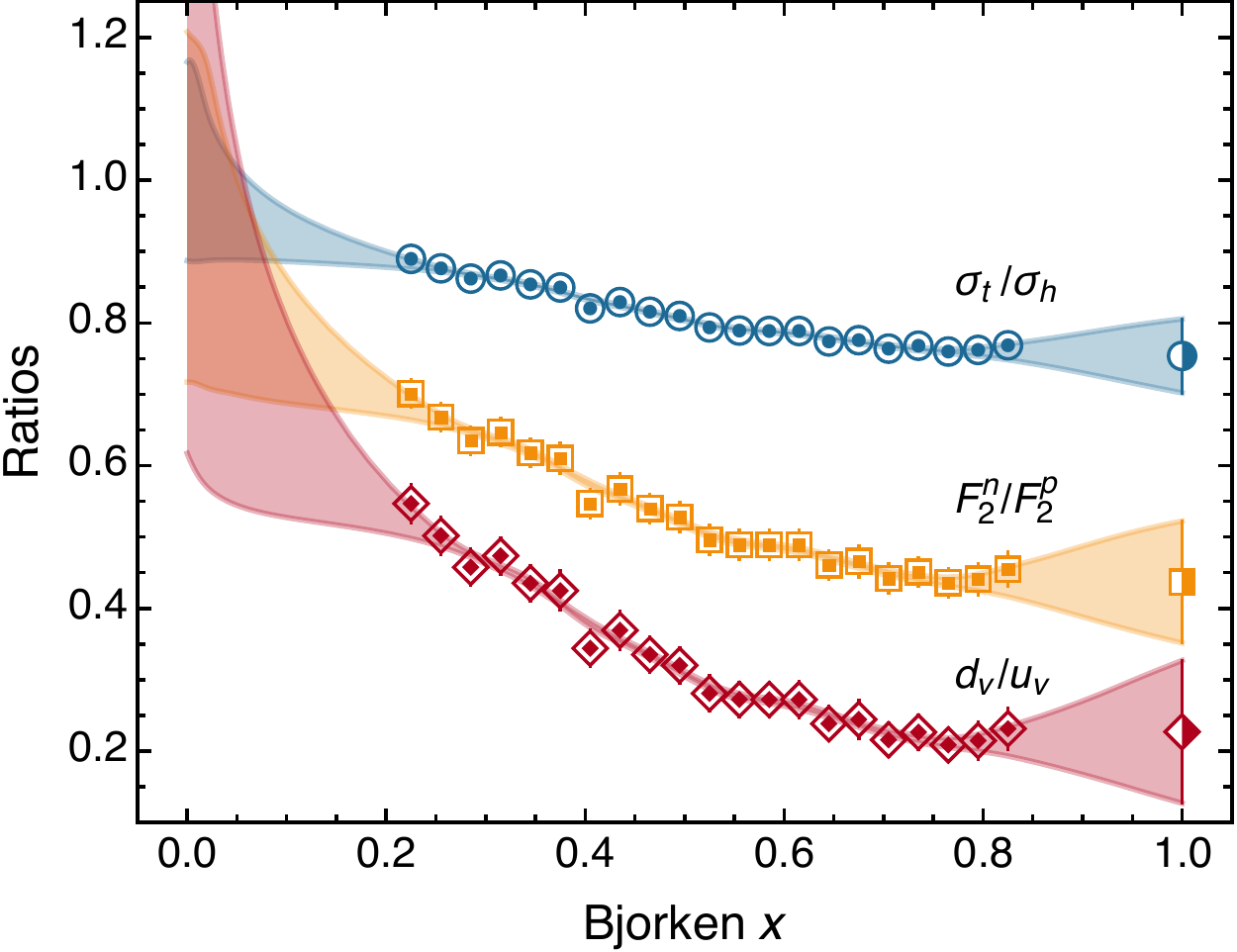}}
\caption{\label{Fmarathon}
\emph{Circles} -- Cross-section ratio $\sigma_t/\sigma_h  = F_2^t/F_2^h$, $h=^3$He, $t=^3$H, with the SPM result drawn as the associated blue band.
\emph{Squares} -- $F^2_n/F^2_p$.  The SPM result is the associated orange band.
\emph{Diamonds} -- $d_v/u_v$ constructed from the $F^2_n/F^2_p$ data using Eq.\,\eqref{F2nF2pV}.  The associated red band is the SPM result obtained from these points.
Data from Ref.\,\cite{Abrams:2021xum}, with $Q^2/$GeV$^2$ increasing from 2.73 to 11.9 as $x$ increases.
}
\end{figure}

The MARATHON data, reproduced in Fig.\,\ref{Fmarathon}, have good coverage of the domain $x\in [0.195,0.825]$ and reach closer to $x=1$ than previous measurements.  Hence, they are well suited for analysis using a newly developed technique to extract an $x=1$ value for $F_2^n(x)/F_2^p(x)$ with a well-defined uncertainty.  The approach, typically described as the statistical Schlessinger point method (SPM) \cite{PhysRev.167.1411, Schlessinger:1966zz}, has been refined in many hadron physics applications, especially those which demand model-independent interpolation and extrapolation \cite{Tripolt:2016cya, Chen:2018nsg, Binosi:2018rht, Binosi:2019ecz, Eichmann:2019dts, Yao:2020vef, Yao:2021pyf, Yao:2021pdy, Cui:2021vgm, Cui:2021aee, Cui:2021skn}.  The SPM avoids any specific choice of fitting function, generating form-unbiased interpolations of data as the basis for well-constrained extrapolations.  Hence, the $x=1$ value produced is model independent, expressing only the information contained in the data.

\medskip

\noindent\emph{4.$\;$Form-unbiased, data-informed extrapolation} ---
The SPM builds a large set of continued fraction interpolations, with each element of the set capturing both local and global features of the curve underlying the data.  The global aspect is vital, guaranteeing the validity of the interpolations outside the data range limits, thereby justifying extrapolation.  The method is detailed elsewhere \cite{Cui:2021vgm}.  Here, therefore, we only recapitulate essentials.

When analysing good experimental data, one must also account for the fact that they are statistically scattered around the curve which truly represents the underlying observable.  The data do not lie on the curve; so, they should not be directly interpolated.  One may address this issue by using the mathematical tool of smoothing with a roughness penalty, following the procedure in Ref.\,\cite{Reinsch:1967aa}.  This approach, sketched in Ref.\,\cite[Sec.\,3]{Cui:2021vgm}, is characterised by an optimal roughness penalty, $\epsilon$, which is a self-consistent output of the procedure.  The data are untouched by smoothing if $\epsilon =0$; whereas maximal smoothing is indicated by $\epsilon =1$.
Regarding the MARATHON data, $\epsilon = 2.0\times 10^{-5} \simeq 0$.

The MARATHON data set has $N=21$ elements; and to proceed with analysis using the statistical SPM, we randomly select
$6 \leq M \leq 15$
elements from the set.  In principle, for each value of $M$, this provides $C(N,M)$ different interpolating functions, which amounts to ${\cal O}(5 \times 10^4 - 3\times 10^5)$ possible interpolators.
For each value of $M$, we then choose the first $1\,000$ curves that are smooth and consistent with Eq.\,\eqref{F2nF2pVNachtmann}.
Every interpolating function defines an extrapolation to any point $x$ outside the range covered by the MARATHON data, including $x=1$.
Thus, for each value of $M$, the value of $F_2^n(x)/F_2^p(x)$ is determined by the average of all results obtained at $x$ from the $1\,000$ curves.

To estimate the error associated with a given SPM-determined value of $F_2^n(x)/F_2^p(x)$, one must first account for the experimental error in the data set.  This is achieved herein by using a statistical bootstrap procedure \cite{10.5555/1403886}: $1\,000$ replicas of the MARATHON data set are generated by replacing each datum by a new point, distributed randomly around a mean defined by the datum itself with variance equal to its associated error.
The fact that $M$ is not fixed introduces a second source of error, $\sigma_{{\delta\!M}}$, whose magnitude we
estimate by shifting $M\to M'$, repeating the aforementioned procedure using $M'$, and evaluating the standard deviation of the distribution of $F_2^n(x)/F_2^p(x)$ for all different $M$ values.

Consequently, the SPM result for $R(x):= F_2^n(x)/F_2^p(x)$ is $R(x) \pm \sigma_{R(x)}$, where:
{\allowdisplaybreaks
\begin{subequations}
\label{SPMR}
\begin{align}
	R(x) & =\sum_{M=6}^{15}\frac{R_M(x)}{10} \,;  \\
\sigma_{R(x)} & =\Bigg[ \sum_{M=6}^{15} \frac{(\sigma^{{M}}_{R(x)})^2}{10^2}
+\sigma_{{\delta\!M}}^2\Bigg]^{\frac12}. \label{sigmadM}
\end{align}
\end{subequations}
At each point $x \in [0,1]$, we have $10$-million results for the ratio, $R(x)$, each calculated from an independent interpolating function.
}

As checks on internal consistency, we also implement this same procedure for the ratio $d_v(x)/u_v(x)$, constructed from the MARATHON data using Eq.\,\eqref{F2nF2pV}, and the primary structure function ratio measured by the Collaboration, \emph{viz}.\ $\sigma_t/\sigma_h=F_2^t/F_2^h$, $h=^3$He, $t=^3$H.

\medskip

\noindent\emph{5.$\;$Validation of SPM extrapolation} ---
Before reporting results, it is first worth demonstrating that the SPM delivers reliable extrapolations when used in connection with the MARATHON data set.  This is achieved following the procedure explained in Ref.\,\cite[Supp.\ Mat.]{Cui:2021vgm}.

We first select a known functional form for the ratio $d_v(x)/u_v(x)$.  Any smooth function consistent with Eq.\,\eqref{F2nF2pVNachtmann} will suffice.  We choose the CJ15 result, determined in Ref.\,\cite{Accardi:2016qay} through a next-to-leading-order global fit to a range of then available data, because that study made a deliberate attempt to reduce uncertainty in $d_v(x)/u_v(x)$ at large-$x$, returning the result
\begin{equation}
\lim_{x\to 1} d_v^{\rm CJ15}(x)/u_v^{\rm CJ15}(x) = 0.09(3)\,.
\end{equation}
Working with the known central function, $R_{15}(x):=d_v^{\rm CJ15}(x)/u_v^{\rm CJ15}(x)$, we generated replica data sets built from values of this ratio at the $x$-points sampled in the MARATHON experiment.  The character of real data is modelled by introducing fluctuations drawn according to a normal distribution.

Treating this new set of data as real, we applied the SPM procedure described above.  Namely:
(\emph{i}) generate $10^3$ replicas;
(\emph{ii}) smooth each replica with the associated optimal parameter;
(\emph{iii}) use the SPM to obtain $R_{15}(1)$ and $\sigma_{R_{15}(1)}$, varying the number of input points $\{M_j=5+j\,\vert\ j = 1,\ldots,10\}$;
and
(\emph{iv}) calculate the final SPM result.
In this way, we obtained
$R_{15}^{\rm SPM}(1) = 0.10(5)$,
\emph{i.e}., reproducing the input value with a 93\% level of confidence.  If one regards the amount of curvature on $x\gtrsim 0.7$ in the CJ15 fits \cite[Fig.\,14]{Accardi:2016qay}, this will be recognised as a keen test and clear validation of the SPM.

We remark that in connection with the extrapolation to $x=1$, there is a discernible sensitivity to the value of $M$, so that $\sigma_{\delta M}$ dominates in Eq.\,\eqref{sigmadM}.  This was not the case when using the SPM to extract the proton charge radius \cite{Cui:2021vgm}; and owes herein to the relative paucity and lower precision of the MARATHON data when compared with that reported for the proton form factor in Ref.\,\cite{Xiong:2019umf}.   This is a mathematical statement, with no other imputations intended: the $\sigma_t/\sigma_h$ measurement is challenging.  Similar observations have been made in connection with the pion charge radius \cite{Cui:2021aee}.

\medskip

\noindent\emph{6.$\;$SPM results from MARATHON} ---
Our SPM results for $\sigma_t/\sigma_h$, $F_2^n/F_2^p$, $d_v/u_v$ are drawn in Fig.\,\ref{Fmarathon}.   Each set of points was analysed independently so that consistency between the various results can serve as additional validations of the method.

Consider first the behaviour on $x\simeq 0$.  Our independent SPM analysis of each data set drawn in Fig.\,\ref{Fmarathon} yields:
\begin{equation}
\label{SPMat0}
\left.
\begin{array}{c}
\sigma_t/\sigma_h \\
F_2^n/F_2^p \\
d_v/u_v
\end{array}\right\}
\stackrel{x\simeq 0}{=}
\left\{
\begin{array}{l}
1.026 \pm 0.139\,, \\
0.962 \pm 0.245\,, \\
1.323 \pm 0.706\,. \\
\end{array}\right.
\end{equation}
Although the statistical uncertainties are significant, especially for $\left. d_v/u_v\right|_{x\simeq 0}$, all results are consistent with sea-quark dominance on $x\simeq 0$; an outcome expected in any experiment consistent with DIS kinematics.
%
%

Turning now to the neighbourhood $x\simeq 1$, the SPM returns the following limiting values:
\begin{equation}
\label{SPMat1}
\left.
\begin{array}{c}
\sigma_t/\sigma_h \\
F_2^n/F_2^p \\
d_v/u_v
\end{array}\right\}
\stackrel{x \simeq 1}{=}
\left\{
\begin{array}{l}
0.754 \pm 0.052\,, \\
0.437 \pm 0.085\,, \\
0.227 \pm 0.100\,. \\
\end{array}\right.
\end{equation}
\emph{Remark 1}.  $\sigma_t/\sigma_h$ data are the source for the $F_2^n/F_2^p$ results.  At each $x$, the relation between them is determined by a single number, ${\cal R}_{ht}$, an EMC ratio \cite{Norton:2003cb, Arrington:2011xs, Hen:2016kwk}, that should be close to unity, \emph{e.g}., \cite[Eq.\,(2)]{Abrams:2021xum}.  Using the results in Eq.\,\eqref{SPMat1}, one finds ${\cal R}_{ht}(x=1)=1.019(13)$.
\emph{Remark 2}.  The $d_v/u_v$ points in Fig.\,\ref{Fmarathon} were obtained from those for $F_2^n/F_2^p$ using Eq.\,\eqref{F2nF2pV}.  The results in Eq.\,\eqref{SPMat1} were obtained via independent SPM treatments of the two data sets.  Using that for $F_2^n/F_2^p$ in Eq.\,\eqref{F2nF2pV}, one finds $d_v/u_v=0.212(10)$, \emph{i.e}., a value just $0.14\,\sigma$ away from that determined using the SPM on the $d_v/u_v$ points.

\medskip

\begin{figure}[t]
\centerline{%
\includegraphics[clip, width=0.42\textwidth]{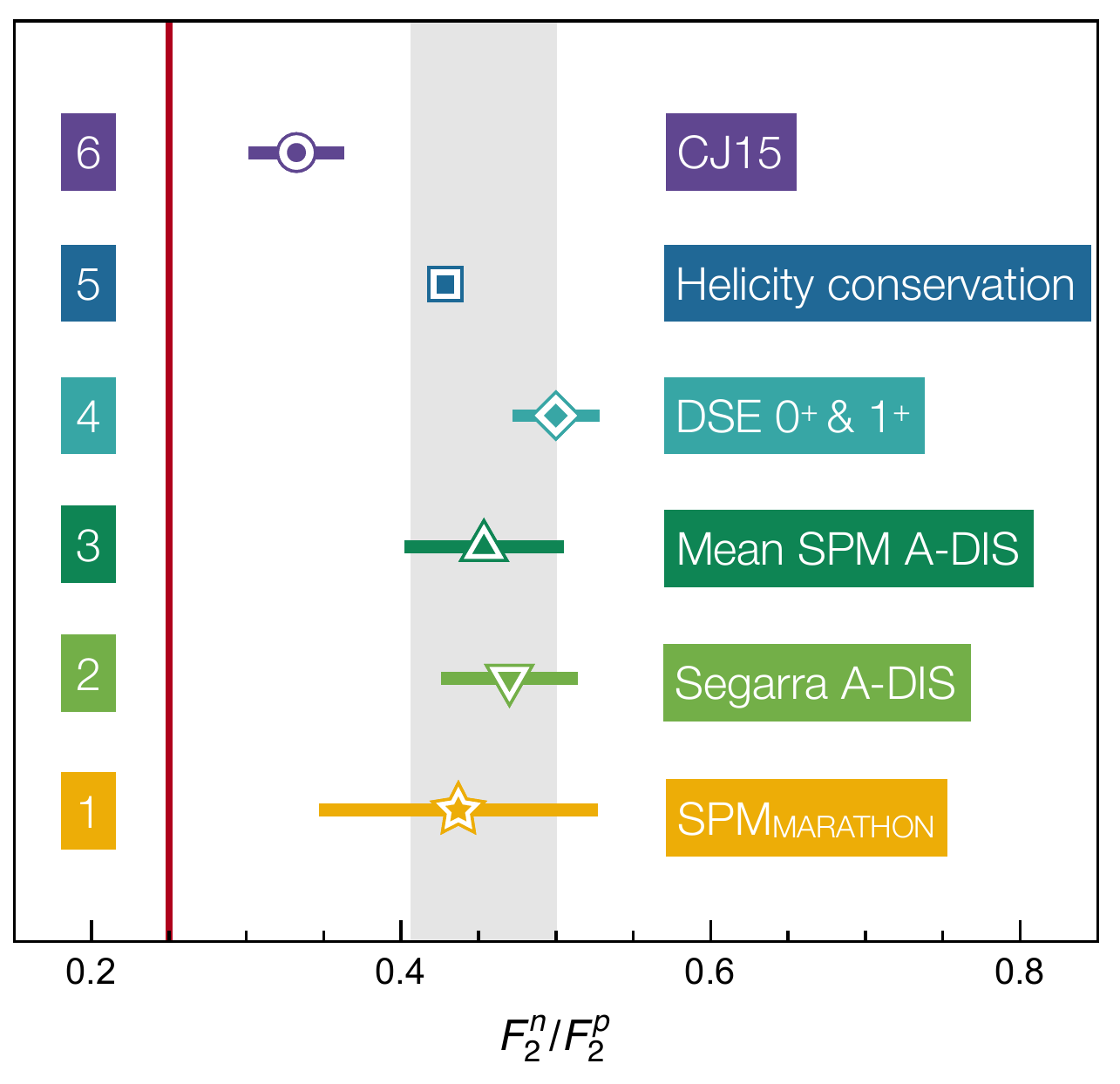}}
\caption{\label{Fconclusion}
$\lim_{x\to 1}F_2^n(x)/F_2^p(x)$.
MARATHON-based SPM prediction compared with results inferred from: nuclear DIS \cite{Segarra:2019gbp};
Dyson-Schwinger equation analyses (DSE) \cite{Segovia:2014aza, Xu:2015kta};
quark counting (helicity conservation) \cite{Farrar:1975yb};
and a phenomenological fit (CJ15) \cite{Accardi:2016qay}.
The vertical red line marks the Nachtmann lower-limit, Eq.\,\eqref{F2nF2pVNachtmann}; 
and row~3 is the average in Eq.\,\eqref{mean}.
}
\end{figure}

\noindent\emph{7.$\;$Conclusions} ---
Fig.\,\ref{Fconclusion} compares our MARATHON-based SPM prediction for $\left. F_2^n/F_2^p\right|_{x\to 1}$ with: the value inferred from nuclear DIS \cite{Segarra:2019gbp}; theory predictions \cite{Segovia:2014aza, Xu:2015kta, Farrar:1975yb, Brodsky:1994kg}; and the result from the phenomenological fit in Ref.\,\cite{Accardi:2016qay}.

The Nachtmann lower bound, Eq.\,\eqref{F2nF2pVNachtmann}, is also highlighted in Fig.\,\ref{Fconclusion}.  It is saturated if valence $d$-quarks play no material role at $x=1$, \emph{i.e}., when there are practically no valence $d$-quarks in the proton: $\left. d_v/u_v\right|_{x\to 1}=0$.  The result is characteristic of models for the proton wave function in which the valence $d$-quark is always sequestered with one of the valence $u$-quarks inside a $J^P=0^+$ ``diquark'' correlation \cite{Close:1988br, Anselmino:1992vg, Barabanov:2020jvn}.  Even allowing for the quark-exchange dynamics typical of Poincar\'e-covariant Faddeev equation treatments of the proton \cite{Barabanov:2020jvn}, one still finds $\left. d_v/u_v\right|_{x\to 1} \approx 0$ if only scalar diquarks are retained \cite{Xu:2015kta}.

\emph{Observation 1}.
Considered alone, the SPM prediction, Eq.\,\eqref{SPMat1}, excludes the result $\left. F_2^n/F_2^p\right|_{x\to 1}=1/4$ with a 99.7\% level of confidence; hence, models of the proton wave function that only include scalar diquarks are excluded with equal likelihood.
Conversely, the QCD parton model prediction \cite{Farrar:1975yb, Brodsky:1994kg}: $d_v(x) \propto u_v(x)$ on $x\simeq 1$, is confirmed with this same level of confidence.

\emph{Observation 2}.
The SPM prediction is consistent with the result inferred from nuclear DIS \cite{Segarra:2019gbp}.  The average of these results is drawn at row 3 in Fig.\,\ref{Fconclusion}:
\begin{equation}
\label{mean}
\left. F_2^n/F_2^p\right|_{x\to 1}^{\rm SPM\, \& \, DIS-A} = 0.454 \pm 0.047\,.
\end{equation}
In this case, the probability that scalar diquark models might be consistent with available data is $0.000014$\%.
%
If Eq.\,\eqref{F2nF2pVNachtmann} is removed as a constraint on SPM interpolants, only the uncertainty in Eq.\,\eqref{mean} changes: $0.047 \to 0.054$.

\emph{Observation 3}.
Eq.\,\eqref{mean} is consistent with both: (\emph{a}) the result obtained by assuming an uncorrelated SU$(4)$ spin-flavour proton wave function and helicity conservation in high-$Q^2$ interactions \cite{Farrar:1975yb, Brodsky:1994kg}; and (\emph{b}) the prediction from a Poincar\'e-covariant Faddeev equation approach to proton structure in which, owing to the mechanisms underling the emergence of hadron mass \cite{Roberts:2020hiw, Roberts:2021nhw}, both $J^P=0^+$ and $J^P=1^+$ diquark correlations are generated with a dynamically determined relative strength \cite{Segovia:2014aza, Xu:2015kta}.

It is here worth noting a strong theory constraint.  Namely, the proton's Poincar\'e covariant bound-state amplitude involves 128 independent Dirac structures, only a small number of which correspond to configurations with a quark+scalar-diquark character \cite{Eichmann:2009qa, Eichmann:2011vu, Qin:2019hgk}.  Consequently, Poincar\'e covariance alone is sufficient to guarantee the presence of additional diquark-like structures in any realistic proton wave function.

In summary, profiting from a combination of today's empirical understanding of target-dependent effects in deep inelastic scattering (DIS), new precise data on the ratio of structure functions in DIS from $^3$He and $^3$H, and a robust method for analysing and extrapolating such data, we arrive at the following result for the proton valence-quark ratio:
\begin{equation}
\label{Final}
\lim_{x\to 1}\frac{d_v(x)}{u_v(x)} = 0.230 \pm 0.057\,.
\end{equation}
With a high level of confidence, proton structure models that are inconsistent with Eq.\,\eqref{Final} can be discarded.

%
\medskip
\noindent\emph{Acknowledgments}.
We are grateful for constructive comments from J.\,R.~Arrington, R.\,W.~Gothe, R.\,J.~Holt, S.-X.~Qin, J.~Rodr\'{\i}guez-Quintero and J.~Segovia.
%
Use of the computer clusters at the Nanjing University Institute for Nonperturbative Physics is gratefully acknowledged.
Work supported by:
National Natural Science Foundation of China (grant nos.\,12135007 and 11805097);
the Alexander von Humboldt Foundation;
and STRONG-2020 ``The strong interaction at the frontier of knowledge: fundamental research and applications'' which received funding from the European Union's Horizon 2020 research and innovation programme (grant no.\,824093).


\begin{thebibliography}{67}
\providecommand{\natexlab}[1]{#1}
\providecommand{\url}[1]{\texttt{#1}}
\providecommand{\urlprefix}{URL }
\expandafter\ifx\csname urlstyle\endcsname\relax
  \providecommand{\doi}[1]{doi:\discretionary{}{}{}#1}\else
  \providecommand{\doi}[1]{doi:\discretionary{}{}{}\begingroup
  \urlstyle{rm}\url{#1}\endgroup}\fi
\providecommand{\bibinfo}[2]{#2}

\bibitem[{Gell-Mann(1964)}]{GellMann:1964nj}
\bibinfo{author}{M.~Gell-Mann}, \bibinfo{title}{{A Schematic Model of Baryons
  and Mesons}}, \bibinfo{journal}{Phys. Lett.} \bibinfo{volume}{8}
  (\bibinfo{year}{1964}) \bibinfo{pages}{214--215}.

\bibitem[{Zweig(1964)}]{Zweig:1981pd}
\bibinfo{author}{G.~Zweig}, \bibinfo{title}{{An SU(3) model for strong
  interaction symmetry and its breaking. Parts 1 and 2}}
  (\bibinfo{year}{1964}) \bibinfo{pages}{(CERN Reports No.\ 8182/TH.\ 401 and
  No.\ 8419/TH.\ 412)}.

\bibitem[{Lucha et~al.(1991)Lucha, Sch{\"o}berl, and Gromes}]{Lucha:1991vn}
\bibinfo{author}{W.~Lucha}, \bibinfo{author}{F.~F. Sch{\"o}berl},
  \bibinfo{author}{D.~Gromes}, \bibinfo{title}{{Bound states of quarks}},
  \bibinfo{journal}{Phys. Rept.} \bibinfo{volume}{200} (\bibinfo{year}{1991})
  \bibinfo{pages}{127--240}.

\bibitem[{Capstick and Roberts(2000)}]{Capstick:2000qj}
\bibinfo{author}{S.~Capstick}, \bibinfo{author}{W.~Roberts},
  \bibinfo{title}{{Quark models of baryon masses and decays}},
  \bibinfo{journal}{Prog. Part. Nucl. Phys.} \bibinfo{volume}{45}
  (\bibinfo{year}{2000}) \bibinfo{pages}{S241--S331}.

\bibitem[{Giannini and Santopinto(2015)}]{Giannini:2015zia}
\bibinfo{author}{M.~M. Giannini}, \bibinfo{author}{E.~Santopinto},
  \bibinfo{title}{{The hypercentral Constituent Quark Model and its application
  to baryon properties}}, \bibinfo{journal}{Chin. J. Phys.}
  \bibinfo{volume}{53} (\bibinfo{year}{2015}) \bibinfo{pages}{020301}.

\bibitem[{Roberts et~al.(2021)Roberts, Richards, Horn, and
  Chang}]{Roberts:2021nhw}
\bibinfo{author}{C.~D. Roberts}, \bibinfo{author}{D.~G. Richards},
  \bibinfo{author}{T.~Horn}, \bibinfo{author}{L.~Chang},
  \bibinfo{title}{{Insights into the emergence of mass from studies of pion and
  kaon structure}}, \bibinfo{journal}{Prog. Part. Nucl. Phys.}
  \bibinfo{volume}{120} (\bibinfo{year}{2021}) \bibinfo{pages}{103883}.

\bibitem[{Brodsky and Lepage(1989)}]{Brodsky:1989pv}
\bibinfo{author}{S.~J. Brodsky}, \bibinfo{author}{G.~P. Lepage},
  \bibinfo{title}{{Exclusive Processes in Quantum Chromodynamics}},
  \bibinfo{journal}{Adv. Ser. Direct. High Energy Phys.} \bibinfo{volume}{5}
  (\bibinfo{year}{1989}) \bibinfo{pages}{93--240}.

\bibitem[{Ellis et~al.(1991)Ellis, Stirling, and Webber}]{Ellis:1991qj}
\bibinfo{author}{R.~K. Ellis}, \bibinfo{author}{W.~J. Stirling},
  \bibinfo{author}{B.~R. Webber}, \bibinfo{title}{{\mbox{$\;$}QCD and collider
  physics}}, \bibinfo{publisher}{Cambridge University Press, Cambridge, UK},
  \bibinfo{year}{1991}.

\bibitem[{Holt and Roberts(2010)}]{Holt:2010vj}
\bibinfo{author}{R.~J. Holt}, \bibinfo{author}{C.~D. Roberts},
  \bibinfo{title}{{Distribution Functions of the Nucleon and Pion in the
  Valence Region}}, \bibinfo{journal}{Rev. Mod. Phys.} \bibinfo{volume}{82}
  (\bibinfo{year}{2010}) \bibinfo{pages}{2991--3044}.

\bibitem[{Rojo et~al.(2015)}]{Rojo:2015acz}
\bibinfo{author}{J.~Rojo}, et~al., \bibinfo{title}{{The PDF4LHC report on PDFs
  and LHC data: Results from Run I and preparation for Run II}},
  \bibinfo{journal}{J. Phys. G} \bibinfo{volume}{42} (\bibinfo{year}{2015})
  \bibinfo{pages}{103103}.

\bibitem[{Peng and Qiu(2016)}]{Peng:2016ebs}
\bibinfo{author}{J.-C. Peng}, \bibinfo{author}{J.-W. Qiu}, \bibinfo{title}{{The
  Drell-Yan Process}}, \bibinfo{journal}{The Universe}
  \bibinfo{volume}{4}~(\bibinfo{number}{3}) (\bibinfo{year}{2016})
  \bibinfo{pages}{34--44}.

\bibitem[{Hen et~al.(2017)Hen, Miller, Piasetzky, and Weinstein}]{Hen:2016kwk}
\bibinfo{author}{O.~Hen}, \bibinfo{author}{G.~A. Miller},
  \bibinfo{author}{E.~Piasetzky}, \bibinfo{author}{L.~B. Weinstein},
  \bibinfo{title}{{Nucleon-Nucleon Correlations, Short-lived Excitations, and
  the Quarks Within}}, \bibinfo{journal}{Rev. Mod. Phys.}
  \bibinfo{volume}{89}~(\bibinfo{number}{4}) (\bibinfo{year}{2017})
  \bibinfo{pages}{045002}.

\bibitem[{Bjorken(1969)}]{Bjorken:1968dy}
\bibinfo{author}{J.~D. Bjorken}, \bibinfo{title}{{Asymptotic Sum Rules at
  Infinite Momentum}}, \bibinfo{journal}{Phys. Rev.} \bibinfo{volume}{179}
  (\bibinfo{year}{1969}) \bibinfo{pages}{1547--1553}.

\bibitem[{Taylor(1991)}]{Taylor:1991ew}
\bibinfo{author}{R.~E. Taylor}, \bibinfo{title}{{Deep inelastic scattering: The
  Early years}}, \bibinfo{journal}{Rev. Mod. Phys.} \bibinfo{volume}{63}
  (\bibinfo{year}{1991}) \bibinfo{pages}{573--595}.

\bibitem[{Friedman(1991)}]{Friedman:1991nq}
\bibinfo{author}{J.~I. Friedman}, \bibinfo{title}{{Deep inelastic scattering:
  Comparisons with the quark model}}, \bibinfo{journal}{Rev. Mod. Phys.}
  \bibinfo{volume}{63} (\bibinfo{year}{1991}) \bibinfo{pages}{615--629}.

\bibitem[{Kendall(1991)}]{Kendall:1991np}
\bibinfo{author}{H.~W. Kendall}, \bibinfo{title}{{Deep inelastic scattering:
  Experiments on the proton and the observation of scaling}},
  \bibinfo{journal}{Rev. Mod. Phys.} \bibinfo{volume}{63}
  (\bibinfo{year}{1991}) \bibinfo{pages}{597--614}.

\bibitem[{Friedman et~al.(1991)Friedman, Kendall, and Taylor}]{Friedman:1991ip}
\bibinfo{author}{J.~I. Friedman}, \bibinfo{author}{H.~W. Kendall},
  \bibinfo{author}{R.~E. Taylor}, \bibinfo{title}{{Deep inelastic scattering:
  Acknowledgements}}, \bibinfo{journal}{Rev. Mod. Phys.} \bibinfo{volume}{63}
  (\bibinfo{year}{1991}) \bibinfo{pages}{629}.

\bibitem[{Roberts et~al.(2013)Roberts, Holt, and Schmidt}]{Roberts:2013mja}
\bibinfo{author}{C.~D. Roberts}, \bibinfo{author}{R.~J. Holt},
  \bibinfo{author}{S.~M. Schmidt}, \bibinfo{title}{{Nucleon spin structure at
  very high $x$}}, \bibinfo{journal}{Phys. Lett. B} \bibinfo{volume}{727}
  (\bibinfo{year}{2013}) \bibinfo{pages}{249--254}.

\bibitem[{Dirac(1949)}]{Dirac:1949cp}
\bibinfo{author}{P.~A.~M. Dirac}, \bibinfo{title}{{Forms of Relativistic
  Dynamics}}, \bibinfo{journal}{Rev. Mod. Phys.} \bibinfo{volume}{21}
  (\bibinfo{year}{1949}) \bibinfo{pages}{392--399}.

\bibitem[{Kogut and Soper(1970)}]{Kogut:1969xa}
\bibinfo{author}{J.~B. Kogut}, \bibinfo{author}{D.~E. Soper},
  \bibinfo{title}{{Quantum Electrodynamics in the Infinite Momentum Frame}},
  \bibinfo{journal}{Phys. Rev. D} \bibinfo{volume}{1} (\bibinfo{year}{1970})
  \bibinfo{pages}{2901--2913}.

\bibitem[{Bjorken et~al.(1971)Bjorken, Kogut, and Soper}]{Bjorken:1970ah}
\bibinfo{author}{J.~D. Bjorken}, \bibinfo{author}{J.~B. Kogut},
  \bibinfo{author}{D.~E. Soper}, \bibinfo{title}{{Quantum Electrodynamics at
  Infinite Momentum: Scattering from an External Field}},
  \bibinfo{journal}{Phys. Rev. D} \bibinfo{volume}{3} (\bibinfo{year}{1971})
  \bibinfo{pages}{1382}.

\bibitem[{Bodek et~al.(1973)Bodek, Breidenbach, Dubin, Elias, Friedman,
  Kendall, Poucher, Riordan, Sogard, and Coward}]{Bodek:1973dy}
\bibinfo{author}{A.~Bodek}, \bibinfo{author}{M.~Breidenbach},
  \bibinfo{author}{D.~L. Dubin}, \bibinfo{author}{J.~E. Elias},
  \bibinfo{author}{J.~I. Friedman}, \bibinfo{author}{H.~W. Kendall},
  \bibinfo{author}{J.~S. Poucher}, \bibinfo{author}{E.~M. Riordan},
  \bibinfo{author}{M.~R. Sogard}, \bibinfo{author}{D.~H. Coward},
  \bibinfo{title}{{Comparisons of Deep Inelastic $e p$ and $e n$
  Cross-Sections}}, \bibinfo{journal}{Phys. Rev. Lett.} \bibinfo{volume}{30}
  (\bibinfo{year}{1973}) \bibinfo{pages}{1087}.

\bibitem[{Poucher et~al.(1974)}]{Poucher:1973rg}
\bibinfo{author}{J.~S. Poucher}, et~al., \bibinfo{title}{{High-Energy
  Single-Arm Inelastic $e p$ and $e d$ Scattering at 6-Degrees and
  10-Degrees}}, \bibinfo{journal}{Phys. Rev. Lett.} \bibinfo{volume}{32}
  (\bibinfo{year}{1974}) \bibinfo{pages}{118}.

\bibitem[{Abrams et~al.(2022)}]{Abrams:2021xum}
\bibinfo{author}{D.~Abrams}, et~al., \bibinfo{title}{{Measurement of the
  Nucleon $F^n_2/F^p_2$ Structure Function Ratio by the Jefferson Lab MARATHON
  Tritium/Helium-3 Deep Inelastic Scattering Experiment -- arXiv:2104.05850
  [hep-ex]$\!$}}, \bibinfo{journal}{Phys. Rev. Lett.}  (\bibinfo{year}{2022})
  \bibinfo{pages}{\emph{in press}}.

\bibitem[{Dove et~al.(2021)}]{SeaQuest:2021zxb}
\bibinfo{author}{J.~Dove}, et~al., \bibinfo{title}{{The asymmetry of antimatter
  in the proton}}, \bibinfo{journal}{Nature}
  \bibinfo{volume}{590}~(\bibinfo{number}{7847}) (\bibinfo{year}{2021})
  \bibinfo{pages}{561--565}.

\bibitem[{Zyla et~al.(2020)}]{Zyla:2020zbs}
\bibinfo{author}{P.~Zyla}, et~al., \bibinfo{title}{{Review of Particle
  Physics}}, \bibinfo{journal}{PTEP} \bibinfo{volume}{2020}
  (\bibinfo{year}{2020}) \bibinfo{pages}{083C01}.

\bibitem[{Chang and Roberts(2021)}]{Chang:2021utv}
\bibinfo{author}{L.~Chang}, \bibinfo{author}{C.~D. Roberts},
  \bibinfo{title}{{Regarding the distribution of glue in the pion}},
  \bibinfo{journal}{Chin. Phys. Lett.}
  \bibinfo{volume}{38}~(\bibinfo{number}{8}) (\bibinfo{year}{2021})
  \bibinfo{pages}{081101}.

\bibitem[{Chang et~al.(2022)Chang, Gao, and Roberts}]{Chang:2022jri}
\bibinfo{author}{L.~Chang}, \bibinfo{author}{F.~Gao}, \bibinfo{author}{C.~D.
  Roberts}, \bibinfo{title}{{Parton distributions of light quarks and
  antiquarks in the proton -- arXiv:2201.07870 [hep-ph]}} .

\bibitem[{Nachtmann(1973)}]{Nachtmann:1973mr}
\bibinfo{author}{O.~Nachtmann}, \bibinfo{title}{{Positivity constraints for
  anomalous dimensions}}, \bibinfo{journal}{Nucl. Phys. B} \bibinfo{volume}{63}
  (\bibinfo{year}{1973}) \bibinfo{pages}{237--247}.

\bibitem[{Cui et~al.(2021{\natexlab{a}})Cui, Ding, Gao, Raya, Binosi, Chang,
  Roberts, Rodr\'\i{}guez-Quintero, and Schmidt}]{Cui:2021sfu}
\bibinfo{author}{Z.~F. Cui}, \bibinfo{author}{M.~Ding},
  \bibinfo{author}{F.~Gao}, \bibinfo{author}{K.~Raya},
  \bibinfo{author}{D.~Binosi}, \bibinfo{author}{L.~Chang},
  \bibinfo{author}{C.~D. Roberts},
  \bibinfo{author}{J.~Rodr\'\i{}guez-Quintero}, \bibinfo{author}{S.~M.
  Schmidt}, \bibinfo{title}{{Higgs modulation of emergent mass as revealed in
  kaon and pion parton distributions}}, \bibinfo{journal}{Eur. Phys. J. A}
  \bibinfo{volume}{57}~(\bibinfo{number}{1})
  (\bibinfo{year}{2021}{\natexlab{a}}) \bibinfo{pages}{5}.

\bibitem[{Cockroft and Walton(1932)}]{Cockroft:1932I}
\bibinfo{author}{J.~D. Cockroft}, \bibinfo{author}{E.~T.~S. Walton},
  \bibinfo{title}{{Experiments with High Velocity Positive Ions. I. Further
  Developments in the Method of Obtaining High Velocity Positive Ions}},
  \bibinfo{journal}{Proc. Roy. Soc. Lond. A} \bibinfo{volume}{136}
  (\bibinfo{year}{1932}) \bibinfo{pages}{619--630}.

\bibitem[{Whitlow et~al.(1992)Whitlow, Riordan, Dasu, Rock, and
  Bodek}]{Whitlow:1991uw}
\bibinfo{author}{L.~W. Whitlow}, \bibinfo{author}{E.~M. Riordan},
  \bibinfo{author}{S.~Dasu}, \bibinfo{author}{S.~Rock},
  \bibinfo{author}{A.~Bodek}, \bibinfo{title}{{Precise measurements of the
  proton and deuteron structure functions from a global analysis of the SLAC
  deep inelastic electron scattering cross-sections}}, \bibinfo{journal}{Phys.
  Lett. B} \bibinfo{volume}{282} (\bibinfo{year}{1992})
  \bibinfo{pages}{475--482}.

\bibitem[{Afnan et~al.(2000)Afnan, Bissey, Gomez, Katramatou, Melnitchouk,
  Petratos, and Thomas}]{Afnan:2000uh}
\bibinfo{author}{I.~R. Afnan}, \bibinfo{author}{F.~R.~P. Bissey},
  \bibinfo{author}{J.~Gomez}, \bibinfo{author}{A.~T. Katramatou},
  \bibinfo{author}{W.~Melnitchouk}, \bibinfo{author}{G.~G. Petratos},
  \bibinfo{author}{A.~W. Thomas}, \bibinfo{title}{{Neutron structure function
  and A = 3 mirror nuclei}}, \bibinfo{journal}{Phys. Lett. B}
  \bibinfo{volume}{493} (\bibinfo{year}{2000}) \bibinfo{pages}{36--42}.

\bibitem[{Pace et~al.(2001)Pace, Salme, Scopetta, and Kievsky}]{Pace:2001cm}
\bibinfo{author}{E.~Pace}, \bibinfo{author}{G.~Salme},
  \bibinfo{author}{S.~Scopetta}, \bibinfo{author}{A.~Kievsky},
  \bibinfo{title}{{Neutron structure function $F_2^n(x)$ from deep inelastic
  electron scattering off few nucleon systems}}, \bibinfo{journal}{Phys. Rev.
  C} \bibinfo{volume}{64} (\bibinfo{year}{2001}) \bibinfo{pages}{055203}.

\bibitem[{Segarra et~al.(2020)Segarra, Schmidt, Kutz, Higinbotham, Piasetzky,
  Strikman, Weinstein, and Hen}]{Segarra:2019gbp}
\bibinfo{author}{E.~Segarra}, \bibinfo{author}{A.~Schmidt},
  \bibinfo{author}{T.~Kutz}, \bibinfo{author}{D.~Higinbotham},
  \bibinfo{author}{E.~Piasetzky}, \bibinfo{author}{M.~Strikman},
  \bibinfo{author}{L.~Weinstein}, \bibinfo{author}{O.~Hen},
  \bibinfo{title}{{Neutron Valence Structure from Nuclear Deep Inelastic
  Scattering}}, \bibinfo{journal}{Phys. Rev. Lett.} \bibinfo{volume}{124}
  (\bibinfo{year}{2020}) \bibinfo{pages}{092002}.

\bibitem[{Segarra et~al.(p ph)}]{Segarra:2021exb}
\bibinfo{author}{E.~P. Segarra}, et~al., \bibinfo{title}{{Nucleon off-shell
  structure and the free neutron valence structure from A=3 inclusive electron
  scattering measurements}}, \bibinfo{journal}{arXiv:2104.07130}
  (\bibinfo{year}{hep-ph}) \bibinfo{pages}{2021}.

\bibitem[{Cocuzza et~al.(2021)Cocuzza, Keppel, Liu, Melnitchouk, Metz, Sato,
  and Thomas}]{Cocuzza:2021rfn}
\bibinfo{author}{C.~Cocuzza}, \bibinfo{author}{C.~E. Keppel},
  \bibinfo{author}{H.~Liu}, \bibinfo{author}{W.~Melnitchouk},
  \bibinfo{author}{A.~Metz}, \bibinfo{author}{N.~Sato}, \bibinfo{author}{A.~W.
  Thomas}, \bibinfo{title}{{Isovector EMC Effect from Global QCD Analysis with
  MARATHON Data}}, \bibinfo{journal}{Phys. Rev. Lett.}
  \bibinfo{volume}{127}~(\bibinfo{number}{24}) (\bibinfo{year}{2021})
  \bibinfo{pages}{242001}.

\bibitem[{Schlessinger(1968)}]{PhysRev.167.1411}
\bibinfo{author}{L.~Schlessinger}, \bibinfo{title}{Use of Analyticity in the
  Calculation of Nonrelativistic Scattering Amplitudes},
  \bibinfo{journal}{Phys. Rev.} \bibinfo{volume}{167} (\bibinfo{year}{1968})
  \bibinfo{pages}{1411--1423}.

\bibitem[{Schlessinger and Schwartz(1966)}]{Schlessinger:1966zz}
\bibinfo{author}{L.~Schlessinger}, \bibinfo{author}{C.~Schwartz},
  \bibinfo{title}{{Analyticity as a Useful Computation Tool}},
  \bibinfo{journal}{Phys. Rev. Lett.} \bibinfo{volume}{16}
  (\bibinfo{year}{1966}) \bibinfo{pages}{1173--1174}.

\bibitem[{Tripolt et~al.(2017)Tripolt, Haritan, Wambach, and
  Moiseyev}]{Tripolt:2016cya}
\bibinfo{author}{R.~A. Tripolt}, \bibinfo{author}{I.~Haritan},
  \bibinfo{author}{J.~Wambach}, \bibinfo{author}{N.~Moiseyev},
  \bibinfo{title}{{Threshold energies and poles for hadron physical problems by
  a model-independent universal algorithm}}, \bibinfo{journal}{Phys. Lett. B}
  \bibinfo{volume}{774} (\bibinfo{year}{2017}) \bibinfo{pages}{411--416}.

\bibitem[{Chen et~al.(2019)Chen, Lu, Binosi, Roberts, Rodr\'\i{}guez-Quintero,
  and Segovia}]{Chen:2018nsg}
\bibinfo{author}{C.~Chen}, \bibinfo{author}{Y.~Lu},
  \bibinfo{author}{D.~Binosi}, \bibinfo{author}{C.~D. Roberts},
  \bibinfo{author}{J.~Rodr\'\i{}guez-Quintero}, \bibinfo{author}{J.~Segovia},
  \bibinfo{title}{{Nucleon-to-Roper electromagnetic transition form factors at
  large $Q^2$}}, \bibinfo{journal}{Phys. Rev. D} \bibinfo{volume}{99}
  (\bibinfo{year}{2019}) \bibinfo{pages}{034013}.

\bibitem[{Binosi et~al.(2019)Binosi, Chang, Ding, Gao, Papavassiliou, and
  Roberts}]{Binosi:2018rht}
\bibinfo{author}{D.~Binosi}, \bibinfo{author}{L.~Chang},
  \bibinfo{author}{M.~Ding}, \bibinfo{author}{F.~Gao},
  \bibinfo{author}{J.~Papavassiliou}, \bibinfo{author}{C.~D. Roberts},
  \bibinfo{title}{{Distribution Amplitudes of Heavy-Light Mesons}},
  \bibinfo{journal}{Phys. Lett. B} \bibinfo{volume}{790} (\bibinfo{year}{2019})
  \bibinfo{pages}{257--262}.

\bibitem[{Binosi and Tripolt(2020)}]{Binosi:2019ecz}
\bibinfo{author}{D.~Binosi}, \bibinfo{author}{R.-A. Tripolt},
  \bibinfo{title}{{Spectral functions of confined particles}},
  \bibinfo{journal}{Phys. Lett. B} \bibinfo{volume}{801} (\bibinfo{year}{2020})
  \bibinfo{pages}{135171}.

\bibitem[{Eichmann et~al.(2019)Eichmann, Duarte, Pe{\~n}a, and
  Stadler}]{Eichmann:2019dts}
\bibinfo{author}{G.~Eichmann}, \bibinfo{author}{P.~Duarte},
  \bibinfo{author}{M.~Pe{\~n}a}, \bibinfo{author}{A.~Stadler},
  \bibinfo{title}{{Scattering amplitudes and contour deformations}},
  \bibinfo{journal}{Phys. Rev. D} \bibinfo{volume}{100} (\bibinfo{year}{2019})
  \bibinfo{pages}{094001}.

\bibitem[{Yao et~al.(2020)Yao, Binosi, Cui, Roberts, Xu, and
  Zong}]{Yao:2020vef}
\bibinfo{author}{Z.-Q. Yao}, \bibinfo{author}{D.~Binosi},
  \bibinfo{author}{Z.-F. Cui}, \bibinfo{author}{C.~D. Roberts},
  \bibinfo{author}{S.-S. Xu}, \bibinfo{author}{H.-S. Zong},
  \bibinfo{title}{{Semileptonic decays of $D_{(s)}$ mesons}},
  \bibinfo{journal}{Phys. Rev. D} \bibinfo{volume}{102} (\bibinfo{year}{2020})
  \bibinfo{pages}{014007}.

\bibitem[{Yao et~al.(2021)Yao, Binosi, Cui, and Roberts}]{Yao:2021pyf}
\bibinfo{author}{Z.-Q. Yao}, \bibinfo{author}{D.~Binosi},
  \bibinfo{author}{Z.-F. Cui}, \bibinfo{author}{C.~D. Roberts},
  \bibinfo{title}{{Semileptonic $B_c \to \eta_c, J/\psi$ transitions}},
  \bibinfo{journal}{Phys. Lett. B} \bibinfo{volume}{818} (\bibinfo{year}{2021})
  \bibinfo{pages}{136344}.

\bibitem[{Yao et~al.(2022)Yao, Binosi, Cui, and Roberts}]{Yao:2021pdy}
\bibinfo{author}{Z.-Q. Yao}, \bibinfo{author}{D.~Binosi},
  \bibinfo{author}{Z.-F. Cui}, \bibinfo{author}{C.~D. Roberts},
  \bibinfo{title}{{Semileptonic transitions: $B_{(s)} \to \pi(K)$; $D_s \to K$;
  $D\to \pi, K$; and $K\to \pi$}}, \bibinfo{journal}{Phys. Lett. B}
  \bibinfo{volume}{824} (\bibinfo{year}{2022}) \bibinfo{pages}{136793}.

\bibitem[{Cui et~al.(2021{\natexlab{b}})Cui, Binosi, Roberts, and
  Schmidt}]{Cui:2021vgm}
\bibinfo{author}{Z.-F. Cui}, \bibinfo{author}{D.~Binosi},
  \bibinfo{author}{C.~D. Roberts}, \bibinfo{author}{S.~M. Schmidt},
  \bibinfo{title}{{Fresh extraction of the proton charge radius from electron
  scattering}}, \bibinfo{journal}{Phys. Rev. Lett.}
  \bibinfo{volume}{127}~(\bibinfo{number}{9})
  (\bibinfo{year}{2021}{\natexlab{b}}) \bibinfo{pages}{092001}.

\bibitem[{Cui et~al.(2021{\natexlab{c}})Cui, Binosi, Roberts, and
  Schmidt}]{Cui:2021aee}
\bibinfo{author}{Z.-F. Cui}, \bibinfo{author}{D.~Binosi},
  \bibinfo{author}{C.~D. Roberts}, \bibinfo{author}{S.~M. Schmidt},
  \bibinfo{title}{{Pion charge radius from pion+electron elastic scattering
  data}}, \bibinfo{journal}{Phys. Lett. B} \bibinfo{volume}{822}
  (\bibinfo{year}{2021}{\natexlab{c}}) \bibinfo{pages}{136631}.

\bibitem[{Cui et~al.(2021{\natexlab{d}})Cui, Binosi, Roberts, and
  Schmidt}]{Cui:2021skn}
\bibinfo{author}{Z.-F. Cui}, \bibinfo{author}{D.~Binosi},
  \bibinfo{author}{C.~D. Roberts}, \bibinfo{author}{S.~M. Schmidt},
  \bibinfo{title}{{Pauli radius of the proton}}, \bibinfo{journal}{Chin. Phys.
  Lett. \emph{Express}} \bibinfo{volume}{38}~(\bibinfo{number}{12})
  (\bibinfo{year}{2021}{\natexlab{d}}) \bibinfo{pages}{121401}.

\bibitem[{Reinsch(1967)}]{Reinsch:1967aa}
\bibinfo{author}{C.~H. Reinsch}, \bibinfo{title}{Smoothing by spline
  functions}, \bibinfo{journal}{Numer. Math.} \bibinfo{volume}{10}
  (\bibinfo{year}{1967}) \bibinfo{pages}{177--183}.

\bibitem[{Press et~al.(2007)Press, Teukolsky, Vetterling, and
  Flannery}]{10.5555/1403886}
\bibinfo{author}{W.~H. Press}, \bibinfo{author}{S.~A. Teukolsky},
  \bibinfo{author}{W.~T. Vetterling}, \bibinfo{author}{B.~P. Flannery},
  \bibinfo{title}{Numerical Recipes 3rd Edition: The Art of Scientific
  Computing}, \bibinfo{publisher}{Cambridge University Press},
  \bibinfo{address}{USA}, \bibinfo{edition}{3} edn., ISBN
  \bibinfo{isbn}{0521880688}, \bibinfo{year}{2007}.

\bibitem[{Accardi et~al.(2016)Accardi, Brady, Melnitchouk, Owens, and
  Sato}]{Accardi:2016qay}
\bibinfo{author}{A.~Accardi}, \bibinfo{author}{L.~T. Brady},
  \bibinfo{author}{W.~Melnitchouk}, \bibinfo{author}{J.~F. Owens},
  \bibinfo{author}{N.~Sato}, \bibinfo{title}{{Constraints on large-$x$ parton
  distributions from new weak boson production and deep-inelastic scattering
  data}}, \bibinfo{journal}{Phys. Rev. D}
  \bibinfo{volume}{93}~(\bibinfo{number}{11}) (\bibinfo{year}{2016})
  \bibinfo{pages}{114017}.

\bibitem[{Xiong et~al.(2019)}]{Xiong:2019umf}
\bibinfo{author}{W.~Xiong}, et~al., \bibinfo{title}{{A small proton charge
  radius from an electron--proton scattering experiment}},
  \bibinfo{journal}{Nature} \bibinfo{volume}{575}~(\bibinfo{number}{7781})
  (\bibinfo{year}{2019}) \bibinfo{pages}{147--150}.

\bibitem[{Norton(2003)}]{Norton:2003cb}
\bibinfo{author}{P.~R. Norton}, \bibinfo{title}{{The EMC effect}},
  \bibinfo{journal}{Rept. Prog. Phys.} \bibinfo{volume}{66}
  (\bibinfo{year}{2003}) \bibinfo{pages}{1253--1297}.

\bibitem[{Arrington et~al.(2012)Arrington, Higinbotham, Rosner, and
  Sargsian}]{Arrington:2011xs}
\bibinfo{author}{J.~Arrington}, \bibinfo{author}{D.~W. Higinbotham},
  \bibinfo{author}{G.~Rosner}, \bibinfo{author}{M.~Sargsian},
  \bibinfo{title}{{Hard probes of short-range nucleon-nucleon correlations}},
  \bibinfo{journal}{Prog. Part. Nucl. Phys.} \bibinfo{volume}{67}
  (\bibinfo{year}{2012}) \bibinfo{pages}{898--938}.

\bibitem[{Segovia et~al.(2014)Segovia, Cloet, Roberts, and
  Schmidt}]{Segovia:2014aza}
\bibinfo{author}{J.~Segovia}, \bibinfo{author}{I.~C. Cloet},
  \bibinfo{author}{C.~D. Roberts}, \bibinfo{author}{S.~M. Schmidt},
  \bibinfo{title}{{Nucleon and $\Delta$ elastic and transition form factors}},
  \bibinfo{journal}{Few Body Syst.} \bibinfo{volume}{55} (\bibinfo{year}{2014})
  \bibinfo{pages}{1185--1222}.

\bibitem[{Xu et~al.(2015)Xu, Chen, Cloet, Roberts, Segovia, and
  Zong}]{Xu:2015kta}
\bibinfo{author}{S.-S. Xu}, \bibinfo{author}{C.~Chen}, \bibinfo{author}{I.~C.
  Cloet}, \bibinfo{author}{C.~D. Roberts}, \bibinfo{author}{J.~Segovia},
  \bibinfo{author}{H.-S. Zong}, \bibinfo{title}{{Contact-interaction Faddeev
  equation and, \emph{inter alia}, proton tensor charges}},
  \bibinfo{journal}{Phys. Rev. D} \bibinfo{volume}{92} (\bibinfo{year}{2015})
  \bibinfo{pages}{114034}.

\bibitem[{Farrar and Jackson(1975)}]{Farrar:1975yb}
\bibinfo{author}{G.~R. Farrar}, \bibinfo{author}{D.~R. Jackson},
  \bibinfo{title}{{Pion and Nucleon Structure Functions Near $x=1$}},
  \bibinfo{journal}{Phys. Rev. Lett.} \bibinfo{volume}{35}
  (\bibinfo{year}{1975}) \bibinfo{pages}{1416}.

\bibitem[{Brodsky et~al.(1995)Brodsky, Burkardt, and Schmidt}]{Brodsky:1994kg}
\bibinfo{author}{S.~J. Brodsky}, \bibinfo{author}{M.~Burkardt},
  \bibinfo{author}{I.~Schmidt}, \bibinfo{title}{{Perturbative QCD constraints
  on the shape of polarized quark and gluon distributions}},
  \bibinfo{journal}{Nucl. Phys. B} \bibinfo{volume}{441} (\bibinfo{year}{1995})
  \bibinfo{pages}{197--214}.

\bibitem[{Close and Thomas(1988)}]{Close:1988br}
\bibinfo{author}{F.~E. Close}, \bibinfo{author}{A.~W. Thomas},
  \bibinfo{title}{{The Spin and Flavor Dependence of Parton Distribution
  Functions}}, \bibinfo{journal}{Phys. Lett. B} \bibinfo{volume}{212}
  (\bibinfo{year}{1988}) \bibinfo{pages}{227}.

\bibitem[{Anselmino et~al.(1993)Anselmino, Predazzi, Ekelin, Fredriksson, and
  Lichtenberg}]{Anselmino:1992vg}
\bibinfo{author}{M.~Anselmino}, \bibinfo{author}{E.~Predazzi},
  \bibinfo{author}{S.~Ekelin}, \bibinfo{author}{S.~Fredriksson},
  \bibinfo{author}{D.~B. Lichtenberg}, \bibinfo{title}{{Diquarks}},
  \bibinfo{journal}{Rev. Mod. Phys.} \bibinfo{volume}{65}
  (\bibinfo{year}{1993}) \bibinfo{pages}{1199--1234}.

\bibitem[{Barabanov et~al.(2021)}]{Barabanov:2020jvn}
\bibinfo{author}{M.~Y. Barabanov}, et~al., \bibinfo{title}{{Diquark
  Correlations in Hadron Physics: Origin, Impact and Evidence}},
  \bibinfo{journal}{Prog. Part. Nucl. Phys.} \bibinfo{volume}{116}
  (\bibinfo{year}{2021}) \bibinfo{pages}{103835}.

\bibitem[{Roberts(2020)}]{Roberts:2020hiw}
\bibinfo{author}{C.~D. Roberts}, \bibinfo{title}{{Empirical Consequences of
  Emergent Mass}}, \bibinfo{journal}{Symmetry} \bibinfo{volume}{12}
  (\bibinfo{year}{2020}) \bibinfo{pages}{1468}.

\bibitem[{Eichmann et~al.(2010)Eichmann, Alkofer, Krassnigg, and
  Nicmorus}]{Eichmann:2009qa}
\bibinfo{author}{G.~Eichmann}, \bibinfo{author}{R.~Alkofer},
  \bibinfo{author}{A.~Krassnigg}, \bibinfo{author}{D.~Nicmorus},
  \bibinfo{title}{{Nucleon mass from a covariant three-quark Faddeev
  equation}}, \bibinfo{journal}{Phys. Rev. Lett.} \bibinfo{volume}{104}
  (\bibinfo{year}{2010}) \bibinfo{pages}{201601}.

\bibitem[{Eichmann(2011)}]{Eichmann:2011vu}
\bibinfo{author}{G.~Eichmann}, \bibinfo{title}{{Nucleon electromagnetic form
  factors from the covariant Faddeev equation}}, \bibinfo{journal}{Phys. Rev.
  D} \bibinfo{volume}{84} (\bibinfo{year}{2011}) \bibinfo{pages}{014014}.

\bibitem[{Qin et~al.(2019)Qin, Roberts, and Schmidt}]{Qin:2019hgk}
\bibinfo{author}{S.-X. Qin}, \bibinfo{author}{C.~D. Roberts},
  \bibinfo{author}{S.~M. Schmidt}, \bibinfo{title}{{Spectrum of light- and
  heavy-baryons}}, \bibinfo{journal}{Few Body Syst.} \bibinfo{volume}{60}
  (\bibinfo{year}{2019}) \bibinfo{pages}{26}.

\end{thebibliography}

\end{document}